\documentclass[aps,prl,twocolumn,showpacs,prlprintnumbers,amsmath,amssymb]{revtex4}
\usepackage{amsmath}
\usepackage{graphicx}
\usepackage{dcolumn}
\usepackage{bm}
\usepackage{overpic}
\usepackage{booktabs}
\usepackage{color}
\usepackage{placeins}
\usepackage{multirow}
\usepackage{xcolor}
\usepackage{multirow}

\newcommand{\el}{El Ni\~{n}o}
\newcommand{\la}{La Ni\~{n}a}

\begin{document}

\title{Climate network suggests enhanced \el ~global impacts in localized areas}
\author{Jingfang Fan$^1$, Jun Meng$^{1,2,3}$\footnote{jun.meng.phy@gmail.com}, Yosef Ashkenazy$^2$\footnote{ashkena@bgu.ac.il}, and Shlomo Havlin$^1$\footnote{havlin@ophir.ph.biu.ac.il}}

\affiliation{1 Department of Physics, Bar-Ilan University, Ramat-Gan 52900, Israel\\
  2 Solar Energy and Environmental Physics, Blaustein Institutes for Desert Research, Ben-Gurion University of the Negev, Israel\\
  3 Laboratory of Soft Matter and Biological Physics, Institute of Physics, Chinese Academy of Sciences,  Beijing, China
 }

\pacs{92.10.am, 05.40.-a, 89.60.-k, 89.75.-k}
\date{\today}

\begin{abstract}
  We construct directed and weighted climate networks based on near surface air temperature to investigate the global impacts of \el ~and \la. We find that regions which are characterized by higher positive/negative network ``in''-weighted links, are exhibiting stronger correlations with the \el ~basin and are warmer/cooler during \el/\la ~periods. These stronger in-weighted activities are found to be concentrated in localized areas, as compared to non-\el ~periods, whereas a large fraction of the globe is not influenced by the events. The regions of localized activity vary from one \el ~(\la) ~event to another; still some \el ~(\la) ~events are more similar to each other. We quantify this similarity using network community structure. The results and methodology reported here may be used to improve the understanding and prediction of \el/\la ~events and also may be applied in the investigation of other climate variables.
\end{abstract}
\maketitle

More than a decade ago, networks became the standard framework with which to study complex systems~\cite{watts1998collective,barabasi1999emergence,cohen2010complex,newman2010networks}.  In recent years, network theory has been implemented in climate sciences to form ``climate networks.'' These have been used successfully to analyze, model, understand, and even predict climate phenomena~\cite{Tsonis2006,Tsonis2007,yamasaki2008climate,Donges2009a,Donges2009b,Steinhaeuser2010, Steinhaeuser2011,Barreiro2011,Deza2013,ludescher2013very,ludescher2014very}.  Specific examples of climate network studies include the investigation of the interaction structure of coupled climate subnetworks~\cite{Donges2011}, the multiscale dependence within and among climate variables~\cite{Steinhaeuser2012}, the temporal evolution and teleconnections of the North Atlantic Oscillation~\cite{Guez2012,Guez2013}, the finding of the dominant imprint of Rossby waves~\cite{wang2013dominant}, the optimal paths of teleconnection~\cite{zhou2015teleconnection}, the influence of \el ~on remote regions~\cite{tsonis2008topology,yamasaki2008climate,gozolchiani2011emergence}, the distinction of different types of \el ~events~\cite{Radebach}, and the prediction of these events~\cite{ludescher2013very,ludescher2014very}.  A network is composed of nodes and links; in a climate network, the nodes are the geographical locations and the links are the correlations between them. The ``strength'' of the links is quantified according to the strength of the correlations between the different nodes~\cite{wang2013dominant,Barrat,Zemp}.

\el ~is probably the strongest climate phenomenon that occurs on decadal time scales. \el ~refers to the warming of the eastern tropical Pacific Ocean by several degrees $^\circ \mathrm{C}$. After an \el ~event, the eastern Pacific Ocean cools by several degrees; this is referred to as \la. This cycle occurs every 3-5 years with different magnitudes. The \el ~phenomenon strongly impacts the local climate and also remote regions including North America~\cite{Ropelewski}, Australia~\cite{Chiew,Power}, Europe~\cite{bronnimann}, the South China Sea, the Indian Ocean, and the tropical North Atlantic~\cite{Klein}. It can lead to warming, enhanced rain in some regions and droughts in other regions, decline in fishery, famine, plagues, political and social unrest, and economic changes. \el ~is a coupled ocean-atmosphere phenomenon which has been linked to internal oceanic Kelvin and Rossby tropical wave activity and to the wind activity above the equatorial Pacific Ocean. There are several indices that quantify the \el ~activity, including the Ni\~{n}o 3.4 index and the Oceanic Ni\~{n}o Index (ONI), which is NOAA's primary indicator for monitoring \el ~and \la.  ONI is the running three-month mean sea surface temperature (SST) anomaly for the Ni\~{n}o 3.4 region (i.e., $5 {^\circ }N- 5 {^\circ }S$, $120 {^\circ }- 170 {^\circ }W$); here we refer to this region as the \el ~Basin (ENB).  When the ONI exceeds $0.5 {^\circ \mathrm{C}}$ for at least five consecutive months, the corresponding year is considered to be an \el ~year. The higher the ONI is, the stronger the \el. Similarly \la ~is determined to occur when the ONI drops below the $-0.5 {^\circ \mathrm{C}}$ anomaly for at least five consecutive months. Presently, we have just undergone one of the strongest \el ~events since 1948~\cite{Levine,Kintisch}. 

Earlier studies used climate networks to investigate the \el ~phenomenon~\cite{ludescher2013very,ludescher2014very,tsonis2008topology,yamasaki2008climate,gozolchiani2011emergence, Radebach}. Most studies defined a weight for each link based on the cross-correlation function and set a threshold with which to identify significant links. Here, we construct a network by using only the ``in''-directed links outgoing from the ENB. The constructed climate network enabled us not only to obtain a map of the global impacts of a given \el, but also to study the local impacts of \el ~in specific regions. These are achieved for the first time by using our new approach. In addition, using only previous events' data, our results confirm most of the regions that were affected during the recently concluded \el ~\cite{Kintisch}.

In the present study, we identify warming and cooling regions which are influenced by the ENB by measuring each node's strength according to the weights of the ``in''-links outgoing from the ENB. We find that during \el/\la, a large fraction of the globe is {\it not} influenced by the events, but the regions that are influenced are significantly more affected by the ENB during \el/\la ~than in normal years.  Our results also indicate that the \el/\la ~events influence different regions with different magnitudes during different events; still by determining the network community structure, our results suggest that similarities exist among some of the \el ~(\la) ~events.

Our evolving climate network is constructed from the global daily near surface ($1000$ hPa) air temperature fields of the National Center for Environmental Prediction/National Center for Atmospheric Research (NCEP/NCAR) reanalysis dataset~\cite{kalnay1996ncep}; see the Supplementary Information for the analysis and results based on the European Centre for Medium-Range Weather Forecasts Interim Reanalysis  (ERA-Interim)~\cite{dee2011era,ERA}. The spatial (zonal and meridional) resolution of the data is $2.5 {^\circ}$ $\times$ $2.5 {^\circ}$, resulting in $144\times 73=10512$ grid points. The dataset spans the time period between January $1948$ and April $2016$~\cite{timeserie}. To avoid the strong effect of seasonality, we subtract the mean seasonal cycle and divide by the seasonal standard deviation for each grid point time series. The network analysis is based on a sequence of networks, each constructed from time series that span one year.
 
\begin{figure}
\begin{centering}
\includegraphics[width=0.48\textwidth]{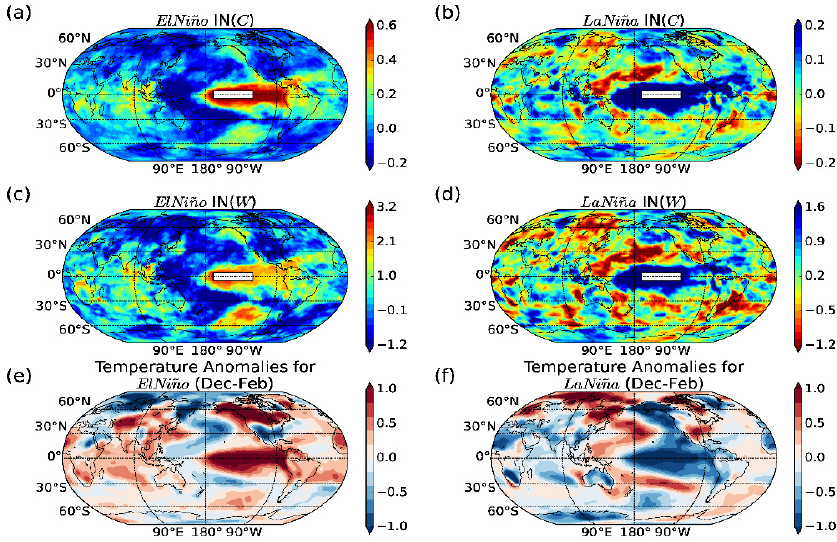}
\caption{\label{Fig:1}(color online). (a), (c) ``In''-weight maps (using $C$ and $W$) for \el ~events. (b),(d), ``In''-weight maps (using $C$ and $W$) for \la ~events. (e), (f)  Mean winter (Dec-Feb) temperature anomalies during \el ~and \la.}
\par\end{centering}
\end{figure}

The nodes (grid points) are divided into two subsets. One subset includes the nodes within the ENB ($57$ nodes) and the other the nodes outside the ENB ($10455$ nodes).  For each pair of nodes, $i$ and $j$, each from a different subset, the cross-correlation between the two time series of $365$ days is calculated, $C_{i,j}^{y}(\tau)$, where $\tau \in [-\tau_{max}, \tau_{max}]$ is the time lag, with $\tau_{max} = 200$ days, and $y$ indicates the starting date of the time series with $0$ time shift. We then identify the value of the highest peak of the absolute value of the cross-correlation function and denote the corresponding time lag of this peak as $\theta^{y}_{i,j}$.  The sign of $\theta^{y}_{i,j}$ indicates the direction of each link---when $\theta^{y}_{i,j} > 0$, the link is regarded as ``outgoing'' from node $i$ and ``incoming'' to node $j$, and vice versa when $\theta^{y}_{i,j} < 0$~\cite{gozolchiani2011emergence}. We only consider links that are ``outgoing'' from the ENB with $|\theta^{y}_{i,j}| \leq 150$ as we focus on the influence of \el ~on the rest of the world on a time scale of a few months.  The link weights are determined using $C_{i,j}^{y}(\theta)$, and we define the strength of the link as
\begin{equation}
W_{i,j}^{y} = \frac{C_{i,j}^{y}(\theta) - {\rm mean}(C_{i,j}^{y}(\tau))}{{\rm std}(C_{i,j}^{y}(\tau))},
\label{eq1}
\end{equation}
where ``mean'' and ``std'' are the mean and standard deviation of the cross-correlation function~\cite{wang2013dominant,zhou2015teleconnection}. We construct networks based on both $C_{i,j}^{y}(\theta)$ and $W_{i,j}^{y}$ and these are consistent with each other. See Fig.~\ref{Fig:1} (a),(c) for \el ~and (b),(d) for \la; see details below.

The adjacency matrix of a climate network is defined as
\begin{equation}
A_{i,j}^{y} = (1 - \delta_{i,j})H(\theta^{y}_{i,j}),
\label{eq2}
\end{equation}
where $H(x)$ is the Heaviside step function for which $H(x\ge 0)=1$ and $H(x< 0)=0$.  The ``in'' and ``out'' degrees of each node are defined as $I_{i}^{y} = \sum_{j} A_{j,i}^{y}$, $O_{i}^{y} = \sum_{j} A_{i,j}^{y}$ respectively, quantifying the number of links into a node or out from a node. 
We define the total ``in'' weights for each node outside the ENB using $C_{j,i}^{y}$ and $W_{j,i}^{y}$ as
\begin{eqnarray}
{\rm IN}(C_{i}^{y}) &=& \sum\limits_{j\in ENB} A_{j,i}^{y} C_{j,i}^{y}(\theta), \nonumber \\
{\rm IN}(W_{i}^{y}) &=& \sum\limits_{j\in ENB} A_{j,i}^{y} W_{j,i}^{y}.
\label{eq3}
\end{eqnarray}
Larger (smaller) positive (negative) values of ${\rm IN}(C_{i}^{y})$ and ${\rm IN}(W_{i}^{y})$ reflect greater warming (cooling) due to the impact of the ENB. If there are no ``in'' links for a node, both the ``in'' degree and ``in'' weights are zero, indicating no impact of ENB.
 
Based on the ONI, we divide the $68$ years into \el, \la, and normal years. For simplicity, we only consider moderate and strong \el/\la ~events with $|{\rm ONI}|> 1{^\circ \mathrm{C}}$. For each event, we consider the time series from July 1 preceding the event to June 30 of the next year, to cover the whole range of one \el/\la ~period~\cite{ELyear, Seager}. Based on this, we consider $11$ \el ~and $9$ \la ~events between the years $1948$ and $2015$. We calculate the ``in''-weighted degree fields for \el ~and \la ~by taking the average of the same type of events using
\begin{eqnarray}
{\rm IN}(C_{i}) &=& {\sum\limits_{y\in EY(LY)}{\rm IN}(C_{i}^{y})}/S,\nonumber \\
{\rm IN}(W_{i}) &=& {\sum\limits_{y\in EY(LY)}{\rm IN}(W_{i}^{y})}/S,
\label{eq4}
\end{eqnarray}
where $S={\sum\limits_{y\in EY(LY)}I_{i}^{y}}$, and ``EY'' and ``LY'' refer to the years in which \el ~and \la ~began.

It is seen that regions affected by \el/\la, either warming or cooling, such as North America~\cite{Cane}, South America~\cite{Grimm}, Europe~\cite{Fraedrich}, India~\cite{Kumar}, South Africa~\cite{Baylis,Anyamba}, and Australia~\cite{Power}, are characterized by relatively high ``in'' weights [Fig.~\ref{Fig:1} (a), (b), (c), (d)] and by high temperature anomalies [Fig.~\ref{Fig:1} (e),(f)]. The maps of temperature anomalies in Fig.~\ref{Fig:1} (e),(f) are obtained by first calculating a three-month (Dec-Feb) mean temperature anomaly for each year, then taking an average of the mean value over all \el/\la ~years. The \el/\la -related ``in''-weighted degree fields are hemispherically symmetric, to some degree,  in accordance with~\cite{Seager,Seager2005}. 

In Table \ref{comparison}, we compare the ``in''-weighted degree maps of \el/\la ~events with the corresponding temperature anomaly maps by evaluating the cross-correlation between each pair of maps shown in Fig.~\ref{Fig:1}. Note that the different grid points are weighted by the cosine of the latitude, to account for the lower weights (due to the smaller area) at the higher latitudes. The cross-correlation values are found to be high, indicating the similarity between the different measures. For more detail, see Table S2 and S3 of the Supplementary Information.

\begin{table}[htbp]
\centering  %
\caption{\label{comparison}Comparison (using cross-correlation) between the ``in''-weighted degree fields and the \el/\la ~mean winter temperature anomaly shown in Fig.~\ref{Fig:1}.}
\begin{tabular}{  c | c | c  }
\hline
  & El Ni\~{n}o & La Ni\~{n}a \\ \hline
$R_{{\rm IN}(C),T}$ & 0.63 & -0.59  \\ 
$R_{{\rm IN}(W),T}$ & 0.54 & -0.54  \\ 
$R_{{\rm IN}(W),{\rm IN}(C)}$ & 0.92 & 0.95  \\ \hline
\end{tabular}
\end{table}

\begin{figure}
\begin{centering}
\includegraphics[width=0.48\textwidth]{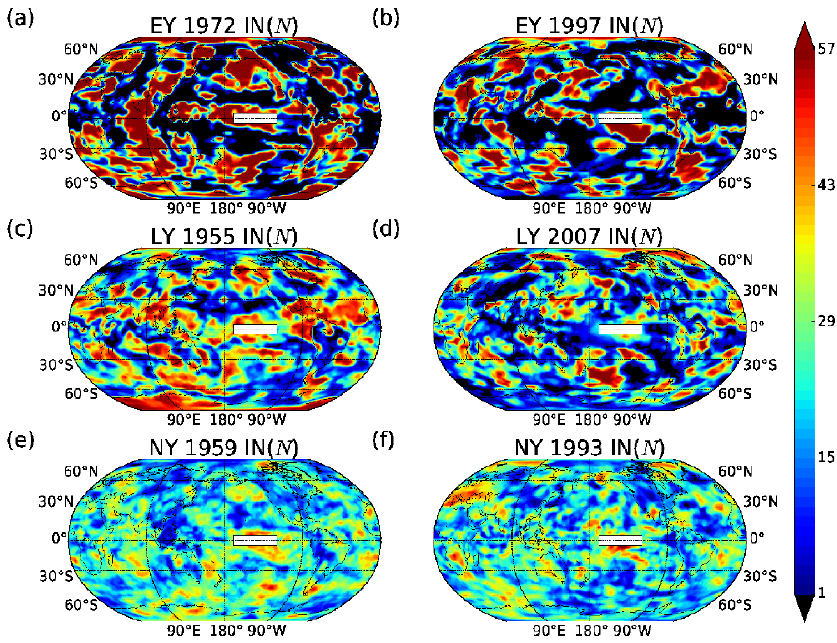}
\caption{\label{Fig:2} (color online) The ``in''-degree fields in typical 
    (a)(b) \el,
    (c)(d) \la,
    and (e)(f) normal years. }
 \label{COMPARE}
\par\end{centering}
\end{figure}

Next we study the variability of the regions that are influenced by \el/\la. We find that during \el/\la ~events, the overall global area that is influenced by these events becomes smaller, while the impact of \el/\la ~in this more limited areas becomes stronger. This is demonstrated in Fig.~\ref{Fig:2}, which compares the global distributions of the ``in'' degrees, $I_{i}^{y}$, of typical \el, ~\la, ~and normal years. The differences are clear---we see broader black regions (that indicate the absence of ``in''-links), as well as broader dark red regions (that indicate that all links connected with the $57$ grid points of the ENB are ``in''-links), during \el ~years [Fig.~\ref{Fig:2} (a),(b)], and, to a lesser degree, during \la ~years [Fig.~\ref{Fig:2} (c),(d)], compared to normal years [Fig.~\ref{Fig:2} (e),(f)]. The underlying reason for this contrast is that during \el/\la, the temperatures of all $57$ nodes located in the ENB are synchronized, such that for each influenced node outside the ENB, the $57$ links connected with the ENB are more likely to have the same direction (i.e., outgoing from the ENB); this situation is less likely during normal years. See also examples of correlations between nodes inside and outside the ENB during \el ~in Fig. S3 of the Supplementary Information.

\begin{figure}
\begin{centering}
\includegraphics[width=0.48\textwidth]{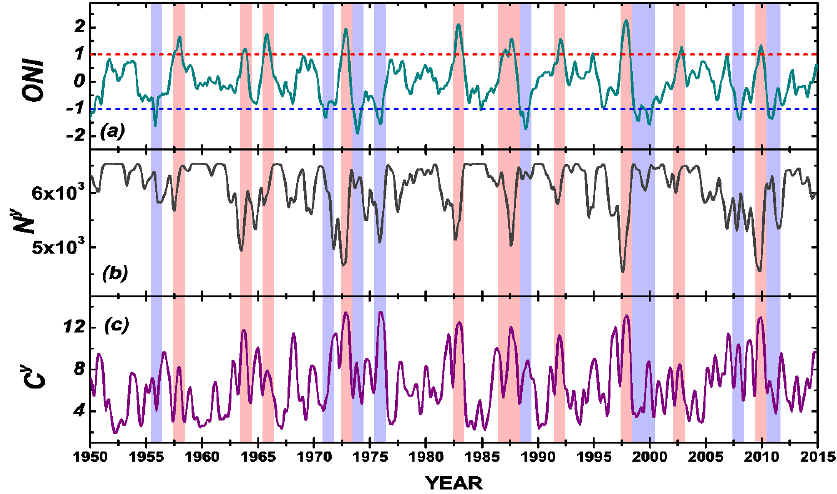}
\caption{\label{Fig:3}(color online)
    (a) The Oceanic Ni\~{n}o Index (ONI) as a function of time. 
    (b) The evolution of the number of nodes that have ``in''-links with time.
    (c) The evolution of the average ``in''-weights per node with time.
     }
 \label{NC}
\par\end{centering}
\end{figure}

A quantitative analysis of the area (number of nodes) that are affected/unaffected during \el ~and \la ~years is shown in Fig.~\ref{Fig:3}, where \el ~and \la ~years are respectively emphasized by the red and blue shading. Here, the temporal evolution of the climate network is studied by constructing a sequence of networks based on successive windows of lengths of $365+200$ days, with a beginning date that is shifted by one month each time. Fig.~\ref{Fig:3} (a) depicts the ONI as a function of time, the dashed horizontal red and blue lines indicating the $\pm1 ^\circ$C values.  We focus on \el~(\la) ~events with ONIs that are larger (smaller) than $1 ^\circ$C ($-1 ^\circ$C).  Fig.~\ref{Fig:3} (b) depicts the number of nodes with no zero ``in''-degree $N^{y}$ as a function of time, and Fig.~\ref{Fig:3} (c) depicts the average ``in'' weights per node, which is given by dividing the sum of the absolute weights of all ``in''-links of each node outside the ENB by $N^{y}$:
\begin{equation}
C^{y}=\sum\limits_{i\not\in ENB}\sum\limits_{j\in ENB} A_{j,i}^{y}\mid C_{j,i}^{y}(\theta)\mid /N^{y}.
\end{equation}
Fig.~\ref{Fig:3} shows the three-month running average of $N^{y}$ and $C^{y}$.

It is seen that during \el/\la, the number of nodes with no ``in''-links, $N^{y}$, drops dramatically [Fig.~\ref{Fig:3}(b)], indicating that the total area influenced by the ENB is smaller. Moreover, during \el/\la ~, $C^{y}$ increases significantly [Fig.~\ref{Fig:3} (c)], indicating a stronger impact of the ENB in the areas that are influenced by it. Other related network quantities are summarized in the Supplementary Information (Figs. S4 to S6 and Table S1). The success of the climate-network-based measures to detect the \el/\la ~events strengthens the reliability of the climate network approach in studying climate phenomenon.

It is possible to classify \el ~events based on the location of their maximum sea surface temperature anomalies and on their tropical mid-latitude teleconnections~\cite{Modoki,Yeh}. Here we propose classifying different types of \el ~events based on the similarity between them, which can be determined by the cross-correlations between pairs of maps (for example, comparing Fig. \ref{Fig:2} (a) and Fig. \ref{Fig:2} (b)) of the ``in''-weighted climate network of different \el ~events. We determine the significance of the cross-correlation using shuffled network maps. The shuffling is performed by dividing the map (globe) into $18$ equal areas, shuffling their spatial orders for each event, and then evaluating the cross-correlation between each pair of the shuffled network global maps. Eventually we obtain a distribution of the cross-correlation values through the shuffling process. Only correlations with $p$ values smaller than $0.01$ are considered as significant.

\begin{figure}
\begin{centering}
\includegraphics[width=0.48\textwidth]{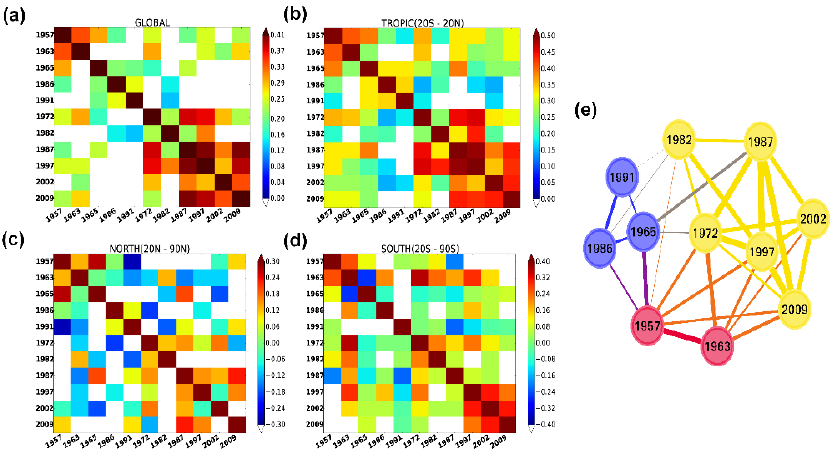}
\caption{\label{Fig:4} (color online) The community structure of the $11$ \el ~events.
    The heat map of cross-correlations between pairs of \el ~events, based on the (a), global, (b) tropical, (c) northern hemisphere, and (d) southern hemisphere maps of the ``in''-weighted climate network. (e) Community structure in the network of $11$ \el ~events. Different colors represent different communities. }
 \label{fig4}
\par\end{centering}
\end{figure}

The cross-correlations between pairs of \el ~events is shown in Fig.~\ref{Fig:4} (a); an insignificant cross-correlation is indicated by the white color. Based on this heat map, the $11$ \el ~events are divided into three groups with extended white areas separating them, indicating that \el ~events within the same group tend to have similar global impact patterns. Furthermore, we divide the globe into three regions, equal in area: ``TROPICS'' ($20^\circ$S to $20^\circ$N), ``NORTH'' ($20^\circ$N to $90^\circ$N), and ``SOUTH'' ($20^\circ$S to $90^\circ$S). Then, separately for each region, we calculate the cross-correlations between the map pairs of the ``in''-weighted climate network. The significant cross-correlations are also determined by $p$ values smaller than $0.01$, by shuffling the spatial orders of nodes within the same regions. The heat maps of cross-correlations for the different regions are shown in Fig.~\ref{Fig:4} (b - d). We find that the global similarity structure receives different contributions from different regions. More specifically, the heat map for the ``TROPICS'' region [Fig.~\ref{Fig:4} (b)] is much more similar to the heat map for the global area [Fig.~\ref{Fig:4} (a)], in comparison to the other two regions, indicating that the impact of \el ~in the tropics dominates the classification of \el ~events. We also construct the matrix of similarity of \el ~events based on the mean winter temperature anomaly and find that it is consistent with the network-based similarity structure; see Fig. S8 of the Supplementary Information. 

A weighted network of the $11$ \el ~years is also constructed based on the significant correlations given in Fig.~\ref{Fig:4} (a) and is shown in Fig.~\ref{Fig:4} (e); the thickness of each link represents the correlation value between the two connected years. Then by utilizing a modularity optimization heuristic algorithm~\cite{Blondel}, our network is sub-divided into three communities, which is consistent with the divided groups in Fig.~\ref{Fig:4} (a).

In summary, a general pattern of \el/\la's global impacts, as well as their dynamical evolutions, are obtained from a time-evolving ``in''-weighted climate network. By averaging the ``in''-weighted degree fields of all significant \el/\la ~events, we identify the regions that tend to be more influenced by \el/\la ~events. The spatial distribution of the ``in''-weighted field of either ${\rm IN}(C)$ or ${\rm IN}(W)$ is consistent with the typical abnormal temperature pattern during \el/\la. Actually, the high consistency between ${\rm IN}(C)$ and ${\rm IN}(W)$ relies on the fact that we only consider links that are ``outgoing'' from the ENB. One of the most important results of our study is that during \el/\la ~periods, a smaller world area is affected by the ENB, but the impact of \el/\la ~is enhanced in these more localized regions. This observation is rooted in the fact that during \el/\la, the entire ENB warms/cools; in addition, the regions that become warmer/cooler have similar/opposite tendencies with the ENB. These synchronized behaviors enhance the overall correlation of the ENB with the rest of the world. However, during normal periods, part of the ENB is correlated and part is not, thus reducing the overall correlation and extending the regions of correlation. Finally, according to our results, different \el ~events can drive different extreme weather conditions in different regions. The recently ended \el ~event was distinct from most \el ~events in certain key aspects of climate disruptions~\cite{Kintisch}. Collecting updated information is important in improving related models and prediction schemes. Meanwhile, the detection of similarities between different \el ~events is also helpful in understanding some common aspects between similar \el ~events.  We distinguish between different types of \el ~events based on the similarities between the networks of these events. According to our results, the similarities between different events are mostly due to the impacts of \el ~in tropical regions.
The methodology and results presented here may facilitate the study of predicting \el ~events and possibly other climate phenomena.

\begin{acknowledgments}
We thank Avi Gozolchiani for helpful discussions.  J.F. Fan thanks the fellowship program funded by the Planning and Budgeting Committee of the Council for Higher Education of Israel. We acknowledge the
MULTIPLEX (No. 317532) EU project, the Israel Science Foundation, ONR and DTRA for financial support.
\end{acknowledgments}

\end{document}